\documentclass[aps,pra,twocolumn,amsfonts, amssymb, amsmath,showpacs,superscriptaddress,showkeys, nofootinbib,longbibliography,]{revtex4-1}
\AtBeginDocument{%
    \newwrite\bibnotes
    \def\bibnotesext{Notes.bib}
    \immediate\openout\bibnotes=\jobname\bibnotesext
    \immediate\write\bibnotes{@CONTROL{REVTEX41Control}}
    \immediate\write\bibnotes{@CONTROL{%
    apsrev41Control,author="08",editor="1",pages="1",title="0",year="1"}}
     \if@filesw
     \immediate\write\@auxout{\string\citation{apsrev41Control}}%
    \fi
}%
\usepackage[T1]{fontenc}
\usepackage[english]{babel}
\usepackage{siunitx}
\usepackage[utf8]{inputenc}
\usepackage{url}
\usepackage{dsfont}
\usepackage{bbm}
\usepackage[usenames,dvipsnames]{xcolor}
\usepackage{nicefrac}
\usepackage{setspace}
\usepackage[short]{ optidef }
\usepackage{float}
\usepackage{comment}

\usepackage[colorlinks=true, citecolor=blue,urlcolor=blue,linkcolor=blue,filecolor=black]{hyperref}

\usepackage{ulem}
\usepackage{amsthm}
\usepackage{mathtools}
\usepackage{physics}
\usepackage{xcolor}
\usepackage{graphicx}
\usepackage[left=23mm,right=13mm,top=35mm,columnsep=15pt]{geometry} 
\usepackage{adjustbox}
\usepackage{placeins}

\usepackage{lipsum}
\usepackage{csquotes}

\newcommand{\beq}{\begin{equation}}
\newcommand{\eeq}{\end{equation}}
\renewcommand{\emph}{\textit}
\renewcommand{\tr}{\text{Tr}}
\usepackage[utf8]{inputenc}
\usepackage{color,soul}

\setul{0.5ex}{0.3ex}
\setulcolor{blue}

\definecolor{pinocol}{rgb}{.1,0.7,0.1}

\definecolor{marcocol}{rgb}{.1,0.2,0.9}
\definecolor{marcocolbg}{rgb}{.1,0.9,0.9}

\begin{document}
\title{Semi-Device-Independent Heterodyne-based Quantum Random Number Generator}

\author{Marco Avesani}
    \affiliation{Dipartimento di Ingegneria dell’Informazione, Università degli Studi di Padova, Padova, Italia}
    
\author{Hamid Tebyanian}
\affiliation{Dipartimento di Ingegneria dell’Informazione, Università degli Studi di Padova, Padova, Italia}
    
\author{Paolo Villoresi}
\affiliation{Dipartimento di Ingegneria dell’Informazione, Università degli Studi di Padova, Padova, Italia}
    
\author{Giuseppe Vallone}
\affiliation{Dipartimento di Ingegneria dell’Informazione, Università degli Studi di Padova, Padova, Italia}
\affiliation{Dipartimento di Fisica e Astronomia, Universit\`a degli Studi di Padova, via Marzolo 8, 35131 Padova, Italy}


\begin{abstract}
Randomness is a fundamental feature of quantum mechanics, which is an invaluable resource for both classical and quantum technologies. Practical quantum random number generators (QRNG) usually need to trust their devices, but their security can be jeopardized in case of imperfections or malicious external actions. In this work, we present a robust implementation of a Semi-Device-Independent QRNG that guarantees both security and fast generation rates.
The system works in a prepare and measure scenario where measurement and source are untrusted, but a bound on the energy of the prepared states is assumed. Our implementation exploits heterodyne detection, which offers increased generation rate and improved long-term stability compared to alternative measurement strategies. In particular, due to the tomographic properties of heterodyne measurement, we can compensate for fast phase fluctuations via post-processing, avoiding complex active phase stabilization systems. As a result, our scheme combines high security and speed with a simple setup featuring only commercial-off-the-shelf components, making it an attractive solution in many practical scenarios. 
\end{abstract}


\maketitle

\section{INTRODUCTION}
Many Quantum Random Number Generators (QRNG) trust the performance of their apparatus, modeling their components as perfect devices \cite{Herrero-Collantes2017, stefanov2000optical, jennewein2000fast}. Although quantum mechanics assures about the random behavior of a quantum process, the practical implementation may be vulnerable to imperfections. Indeed, the trust on the devices may allow information leakage, if the devices are correlated with the environment under the control of an adversary.

Therefore, the security of a practical QRNG is linked to the number of required assumptions, where fewer assumptions imply better security~\cite{Acin2016}. From this point of view, Device-Independent (DI) protocols~\cite{colbeck2011,Acin2016} offer the highest level of security. In this scenario, the instruments are treated as black boxes, and nothing is assumed about their inner working principles. However, all DI protocols require the loophole-free violation of a Bell inequality, which is extremely demanding from an experimental point of view.
Moreover, despite the advancements presented in recent demonstrations ~\cite{Bierhorst2017,Liu2018a,Liu2019b,Zhang2020,Shalm2019}, the generation rate of DI protocol is orders of magnitude lower than what is required by practical applications.

Novel scenarios are required to meet the needs of speed and security. In order to achieve a better trade-off between generation rate and security an alternative approach, called Semi-DI, has been recently proposed~\cite{Ma2016}. 
The Semi-DI approach, by including some assumptions on the working principle of the devices without requiring their full characterization, offers a solution to overcome experimental complexity and low generation-rate issues of DI protocols.
Few different Semi-DI protocols have been proposed that require a trusted  source \cite{Nie2016, Cao2015} or measurement \cite{Vallone2014a, Avesani2018, Marangon}. Other protocols do not require to trust specific components of the setup but they make assumptions on the overlap~\cite{Brask2017} or the energy~\cite{Himbeeck2017semidevice,DavideR,Himbeeck2019CorrelationsConstraints} of the emitted states
or assumptions on the dimension of the Hilbert space~\cite{Lunghi2014b, Canas2014}.

However, the performances offered by Semi-DI protocols are dramatically higher than the DI ones, matching, and sometimes surpassing, the generation rates of commercial QRNG~\cite{Ma2016}. 
In particular, continuous variables (CV) implementations are an attractive solution in this context with respect to discrete variable (DV) ones since they can exploit a larger Hilbert space, fast detectors and they only require standard commercial-off-the-shelves (COTS) components typical of the telecom market. As a result, they feature higher generation rates with simpler optical setups.

In the present work, we introduce a Semi-DI QRNG based on heterodyne detection that assumes an upper bound on the energy of the emitted states. Our implementation ensures an excellent generation rate on par with commercial solutions and improved security.
Additionally, this scheme is realized with an easy-to-implement all-fiber setup. The scheme is based on a prepare-and-measure scenario; therefore, no entanglement is necessary. The assumption required for the execution of the protocol is related to the energy of the states emitted from the source, which is monitored in real-time during the execution of the protocol.

The tomographic capabilities of the heterodyne detection allow us to sample the full phase space, enabling us to track and compensate fluctuations and drifts of the signal phase via software in post-processing, without any active stabilization system. 
This feature makes our implementation more practical with respect to other alternative measurement strategies, where active real-time phase stabilization is required.

The amount of secure and certified randomness is obtained by numerically bounding the quantum conditional min-entropy via Semidefinite programming
(SDP), with an approach similar to the one described in~\cite{Brask2017}: we note that the energy assumption is required for every round of the protocol.
Compared to~\cite{Brask2017} our solution is based on a bound on the energy of the prepared states, rather than the overlap, since the first can be easily monitored experimentally. Moreover, it has a substantially higher generation rate, due to the exploitation of high-speed balanced detectors used in CV measurements rather than slow single-photon detectors.

We note that a similar approach, based on a different measurement apparatus, has been proposed independently in \cite{rusca2020}.

\section{THE PROTOCOL}
\subsection{Semi-DI QRNG based on overlap bound}
\begin{figure}[t]
\centering
\includegraphics[height=2 in]{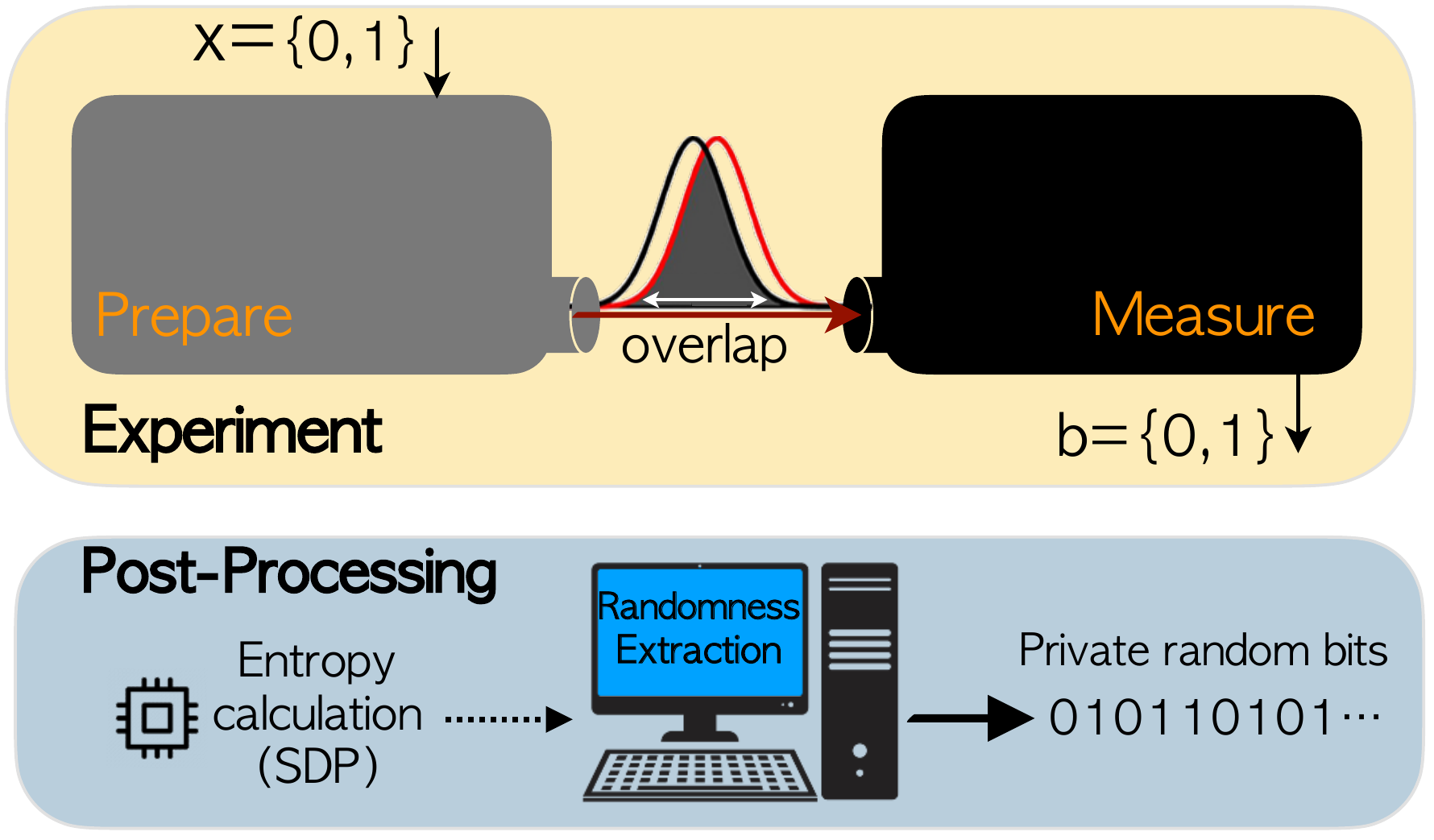}
\caption{General idea of the Semi-DI QRNG protocol.}
\label{fig:setup}
\end{figure}

The general scheme of our QRNG is illustrated in Fig.~\ref{fig:setup}.
In the first step, the preparation box, after taking a binary input $ x \in \{0,1 \}$, 
emits a quantum state $ \rho_x $ which is sent to the measurement box  that performs a measurement with binary outputs $ b \in \{0,1 \}$.

It is worth noticing that the preparation and measurement devices are considered as untrusted  with a single assumption: the energy of the prepared states is upper bounded, as expressed by the following relation
\beq
\label{eq:energy_bound}
\langle \hat n\rangle_{\rho_x}\leq \mu\,.
\eeq
where $\hat n$ is the photon number operator.
As obtained in~\cite{Himbeeck2017semidevice}, when $\mu\leq 0.5$, the upper bound on the energy implies a lower on the overlap between the two states. From the bound on the overlap, it is possible to follow the approach of ~\cite{Brask2017} to obtain the amount of certified private randomness.
Indeed, from the obtained conditional probability outcomes $p(b|x)$ it is possible to compute the conditional min-entropy by a SDP, which estimates how much private random bits are available out of a generated bit strings. A randomness extractor~\cite{Tomamichel2011}, based on the evaluation on the conditional min-entropy $H_{\rm min}$, reduces the string of generated raw bits to a shorter one, which is private and genuinely random.

We now briefly review how the min-entropy can be estimated.
With complete generality, any preparation device will generate the pure quantum state $\ket{\psi^\lambda_x}$ with probability $p(\lambda)$ for each input $x$. The constraint~\eqref{eq:energy_bound} implies a lower bound on the state overlap, namely~\cite{Himbeeck2017semidevice}.
\beq 
\label{eq:overlap_bound}
|\langle\psi^\lambda_0|\psi^{\lambda'}_1\rangle|\geq 1-2\mu
\eeq
At the same time, the measurement device can be modeled as a binary outcome positive-operator valued measurement $\{\Pi _b^\lambda\}$ (POVM) with
$\Pi_0^\lambda+\Pi_1^\lambda=\openone$.
The variable $\lambda$ represents  classical shared randomness between preparation and measurement devices, and correspond to the ``strategy'' of an adversary that 
builds the devices is pursuing in order to guess the output bits $b$. 
Each strategy $\lambda$ is associated with a corresponding probability $p(\lambda)$.
The states $\ket{\psi^\lambda_x}$ and the POVM 
$\Pi _b^\lambda$ could be arbitrary, but they are constrained by Eq.~\eqref{eq:overlap_bound} and by the measured conditional probabilities, namely
\beq
\label{eq:constraints}
\begin{aligned}
&p(b|x) = \sum\limits_\lambda  {p(\lambda )} 
\bra{\psi^\lambda_x}\Pi _b^\lambda\ket{\psi^\lambda_x}
\end{aligned}
\eeq

Given the measured probabilities $p(b|x)$ and the assumed bound on the energy $\mu$, the guessing probability of an adversary $P_g$
averaged over the preparation probability $p_x$ is given by the following relation:
\beq
P_g = \sum_{x,\lambda}p_x p(\lambda) 
\max \left\{ \bra{\psi^\lambda_x}\Pi _0^\lambda\ket{\psi^\lambda_x}
, \bra{\psi^\lambda_x}\Pi _1^\lambda\ket{\psi^\lambda_x}
 ]
 \right\} 
\eeq

From the above equation, $P_g$ is the maximum guessing probability of the outcome averaged over $x$ and $\lambda$.
The fraction of private random bits that can be extracted by the raw output sequence is given by the min-entropy
\beq
\label{eq:Hmin}
H_{\rm min}=-\log_2\left(\max_{\{\ket{\psi^\lambda_x},\Pi^\lambda_b\}}P_g\right)\,,
\eeq
where the maximization is performed over all possible preparation and measurement strategies that satisfy the constraints given in Eqs.~\eqref{eq:overlap_bound} and~\eqref{eq:constraints}.
The upper bound on $P_g$ can be efficiently solved through semidefinite programming (SDP), as reported in~\cite{Brask2017} and reviewed in Appendix~\ref{sec:SDI}. 
In particular, we employ the dual formulation of the SDP, since it provides few advantages with respect to the primal. First, since it involves a maximization problem it always returns a lower-bound on the min-entropy, thus never overestimating it. Secondly, it is less computational demanding.
Indeed, since the objective function is linear in $p(b|x)$, after an optimal solution has been found, a new (non-tight) lower bound with different $p(b|x)$ can be easily evaluated, without running again the optimization. Finally, with the dual formulation finite-size effects due to the limited statistics can be easily taken into account.

We point out that it is also possible to derive a different bound on the min-entropy by constraining the energy of the emitted states, using the framework described in \cite{Himbeeck2017semidevice,Himbeeck2019CorrelationsConstraints}. This scenario has been experimentally implemented using single photon detectors in \cite{DavideR} and homodyne measurement in a recent parallel work.

The number of random bits that can be generated, expressed by Eq.~\eqref{eq:Hmin}
is guaranteed by the laws of quantum mechanics. 
In order to intuitively explain why randomness can be generated, it is possible to consider the measurement as a device that should discriminate between the two input states, with
$p(1|0)$ and $p(0|1)$ the ``error probabilities''.
When the energy is upper bounded and thus the overlap of two states is lower bounded, there is no binary outcome measurement able to perfectly distinguish them, making the probabilities $p(1|0)$ and $p(0|1)$ both vanishing.
Indeed, the minimal error chance $p(1|0)+p(0|1)$ is bounded by the state overlap,
namely 
\beq
\\
p(1|0)+p(0|1)\geq1-2\sqrt{\mu-\mu^2}
\eeq
Thus a nonvanishing overlap  implies a nonvanishing error rate that can be employed as a randomness source. 
\begin{figure}[t!]
\centering
\includegraphics[width=\linewidth]{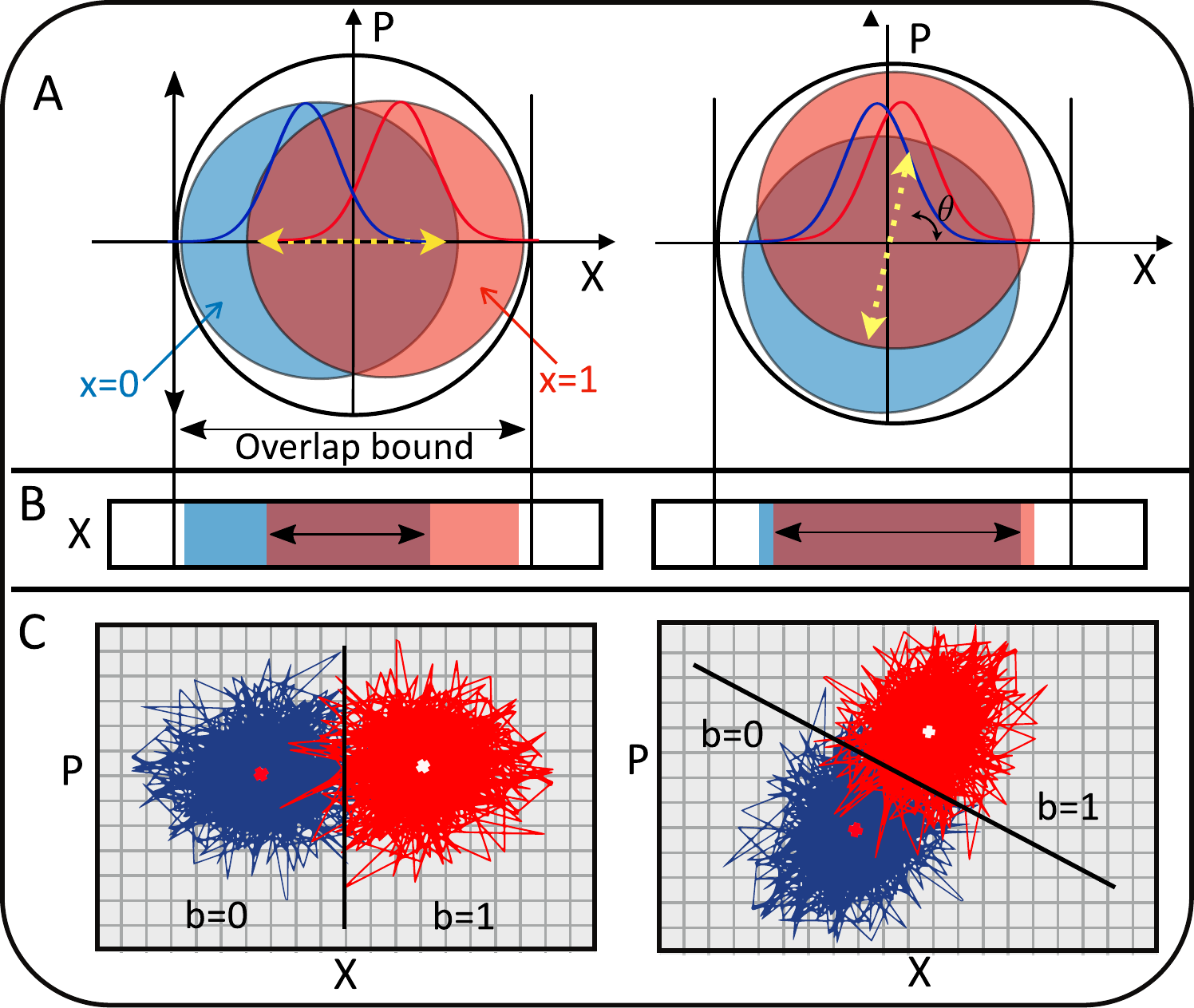}
\caption{A) For $x=0$ and $x=1$, the blue and red states are prepared, respectively. In the left panel  states with phase  $\phi=0$ are prepared; on the right panel, the prepared states have phase $\phi=\theta$.
 B) homodyne detection along $X$ quadrature. When $\theta\neq 0$ the two states become less distinguishable C) Heterodyne measurement: for a binary outcome, if the received state is located on the left side with respect to the linear classifier then $b=0$, otherwise $b=1$. The state distinguishability does not vary with $\theta$ .
}
\label{fig:overlap}
\end{figure}
We stress here that in this protocol no assumption is made on the performance of the apparatus; therefore, possible noise due to unavoidable technical imperfections are already included in the above analysis. 

Notably, neither the selection of the states nor the measurements are known, and randomness certification is only based on the input-output correlations $p(b|x)$, and the upper bound $\mu$ on the energy. 

\subsection{Implementation with heterodyne}
In our experimental implementation, as  displayed in Fig.~\ref{fig:overlap}(A), the source generates the coherent states 
 $ \ket{\psi_0}=\ket{\alpha}$  and
 $ \ket{\psi_1}=\ket{-\alpha}$, with $\ket{\pm\alpha}=
 {e^{-\frac{\mu}{2}}}\sum\limits_{n = 0}^\infty  \frac{(\pm\sqrt{\mu}e^{i\phi})^n}{{\sqrt{n!}}}\ket n$, where $\alpha=\sqrt{\mu}e^{i\phi}$, $\mu$ is the mean photon number and $\phi$ is the relative phase between the signal and LO.
Depending on the input $x$, the phase of a coherent state (produced by a CW laser) is modulated such that the output phase for $x = 0$ is $\phi$, while for $x = 1$ it is $\pi+\phi$. 
 We note that  $\mu$ is precisely the upper bound \eqref{eq:energy_bound} on the state energy.

\begin{figure*}[htb!]
  \includegraphics[width=0.99\linewidth]{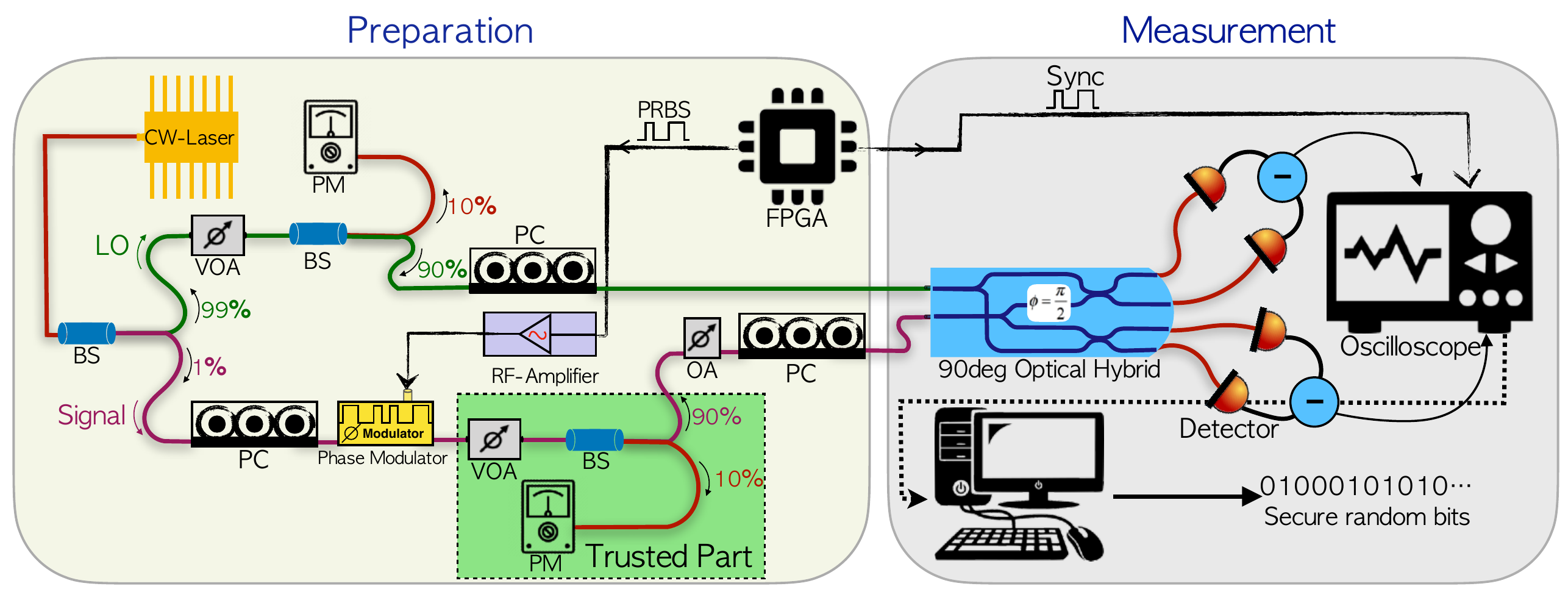}
\caption{Experimental setup, which is consist of two sections, preparation, and measurement. A coherent state is generated by a CW-laser and sent to the interferometer. One arm, with $1\%$ of the light (purple path), is employed to prepare the signal, and the other one with $99\%$ of the light (green route), is the local oscillator. In each path, $10\%$ of light is transmitted to PM for monitoring the power. The two paths are combined on the $90^{\circ}$ optical hybrid, which is followed by a pair of balanced detectors implementing the heterodyne measurement. An FPGA controls the phase modulator and the synchronization with the oscilloscope.}
  \label{SETUP}
\end{figure*}

We underline that the input $x$ must be independent of the devices and in particular, it should be uncorrelated with $\lambda$. In our experiment, $x$ will be generated from a classical RNG (e.g., Pseudo RNG).
The receiver is represented by a heterodyne detector that can be modeled by the POVM 
$\Pi_\beta = \frac{1}{\pi}|\beta\rangle \langle\beta|$. The transmitted state is indeed measured by interfering it with a local oscillator on a $90^{\circ}$ hybrid, followed by balanced detectors in the output ports.

Considering that the heterodyne detection provides information on both field quadratures $X$ and $P$,  any measurement outcome can be represented by a complex number $\beta$ and associated to a point in the $(X,P)$ phase space. These points can be then gathered  in two sets, corresponding to $x=0$ and $x=1$ respectively, as it is shown in Fig.~\ref{fig:overlap}(C). Next, we determine the centroids $C_0$ and $C_1$ of each of these two sets.
The axis of the segment $C_0C_1$ divides the phase space into two regions: each heterodyne measurement $\beta$ is associated to the $b=0$ or $b=1$ output if  $\beta$ belongs to the region containing $C_0$ or $C_1$ respectively.
From the heterodyne detection POVM $\Pi_\beta$, it is easy to compute the theoretical output probabilities as
\beq
\begin{aligned}
\label{eq:conditional_prob}
p(0|0) &= p(1|1)= \frac{1}{2}(1 + \erf(|\alpha|))
\\
p(0|1) &= p(1|0) = \frac{1}{2}(1 - \erf(|\alpha|))
\\
\end{aligned}
\eeq
with $\erf(x)=\frac{2}{\sqrt{\pi}}\int^{x}_0e^{-t^2}\text{d}t$ the error function.

Given these probabilities, we can use the SDP to calculate the expected performances of an honest implementation with ideal devices,  as a function of $\alpha$. These expectations are plotted, together with  the experimental results, in Fig. \ref{fig:res}.

The main advantage of heterodyne detection over alternative measurement strategies is given by its intrinsic robustness to phase drifts. In practical implementations, the relative phase $\phi$ between the signal and the LO, rapidly drifts over time, due to thermal or mechanical fluctuations. As a result, the measured states rapidly rotate around the center of the phase space (see Fig. \ref{fig:overlap} A). For measurements that are sensitive only to one quadrature, such as homodyne detection, any deviation from the condition $\phi=0,\pi$ reduces the distinguishability between the two states $\ket{\psi_0},\ket{\psi_1}$ reducing the extractable randomness (see Fig. \ref{fig:overlap} B). In the extreme case where $\phi=\frac\pi2,\frac{3\pi}{2}$, the two states are indistinguishable meaning that no randomness can be certified at all. Thus, for these systems is mandatory to implement a fast and active phase stabilization system, running in real-time, that is able to  keep the relative phase $\phi$ fixed at the optimal point. The active stabilization is not trivial to realize and substantially increases the experimental complexity of the system. 
On the contrary, heterodyne detection, being tomographically complete, is able to sample the full phase-space and a drift of the relative phase $\phi$ doesn't affect the distinguishability of the states $\ket{\psi_0},\ket{\psi_1}$  and it is equivalent to a rotation of the reference system (see Fig. \ref{fig:overlap} C). So, if the sampling rate of the system is sufficiently fast with respect to the time-scale of the fluctuations, the relative phase $\phi$ can be tracked \textit{a posteriori} via software. With this solution it is possible to overcome much faster drifts with drastically simpler experimental setups.
In Appendix~\ref{sec:homo}, 
we describe with more details the differences between heterodyne and homodyne detection schemes.

\section{Experiment}
\subsection{The Optical and Electronic Part}
The experimental setup 
that implements the scheme represented in Fig.~\ref{fig:setup} is drawn in Fig. \ref{SETUP}.
A continuous mode laser generates a coherent state at 1550 nm. A $99:1$ beam spliter (BS), is used to split the light into two branches, local oscillator (LO), and signal. The LO is transferred to an automatic Variable Optical Attenuator (VOA) and then is divided again with a $90:10$ BS.
$10\%$ of the light is sent to a power meter (PM) for calibrating the detectors. It should be noted that there is no assumption on the measurement device, and this calibration is done in order to monitor the correct working of the detectors, but is not strictly required and it does not influence the security of the protocol.
The remaining $90\%$ of the light is sent to a fiber Polarization Controller (PC) and then to the LO port of the $90^{\circ}$ optical hybrid. 
The optical signal used to prepare the states $|\psi_x\rangle$, is transmitted to a PC followed by a fiber LiNbO3 phase modulator (MPZ-LN-20 by iXblue) with a bandwidth of $20\si{\giga\hertz}$. A Pseudo-Random Binary Sequence (PRBS) $x$ is generated in real-time by a Field Programmable Gate Array (FPGA) with a rate of $1.25\si{\giga\hertz}$. The output of the FPGA is amplified with an RF amplifier (iXblue) and then used to drive the phase modulators, adding $0$ or $\pi$ phase shift to the light inside the phase modulator, thus preparing the states $\ket{\psi_0}=\ket{\alpha}$ or $\ket{\psi_1}=\ket{-\alpha}$ respectively.

The modulated light is then conveyed to a mechanical VOA, used to change the magnitude of $\alpha$ before being split by a $90:10$ BS. $90\%$ of the light is sent to a PM (Thorlabs S154Cwith a measurement uncertainty of $\pm5\%$), while $10\%$ is sent to a fixed optical attenuator (OA).  With this configuration, after calibrating the attenuation entered by the OA, we can have a one-to-one mapping between the power read on the PM and the optical power sent to the measurement part. For the estimation of the mean photon number $\mu$ we consider a time slot of $0.8 \si{\nano\second}$, determined by the system repetition rate. Before entering the signal port of the  $90^{\circ}$ optical hybrid, the polarization of the optical signal is adjusted with a PC.

After the optical hybrid, the two pais of optical signals relative to the two quadratures are sent to two InGaS Balanced Photo-receiver (PDB480C-AC) with a bandwith of $1.6\si{\giga\hertz}$. At the receiver side, the RF signal generated by the balanced photodetectors and a synchronization signal coming from the FPGA are digitized by Tektronix DPO70004 Oscilloscope with $4\si{\giga\hertz}$ of analog bandwidth, at a sampling rate of $12.5\si{\giga}\text{sps}$ and $8$ bit of resolution. We average every 10 samples in order to obtain signals with the same repetition rate of the source. This oversampling procedure is used to better reconstruct the signal from the balanced detectors, that shows a finite rise-time and electronic ringing, due to the high repetition rate of the system. The oscilloscope works in burst mode, saving the acquired data in its internal storage memory. The post-processing of the data is done offline.

\begin{figure}[t]
\centering
\includegraphics[width=\linewidth]{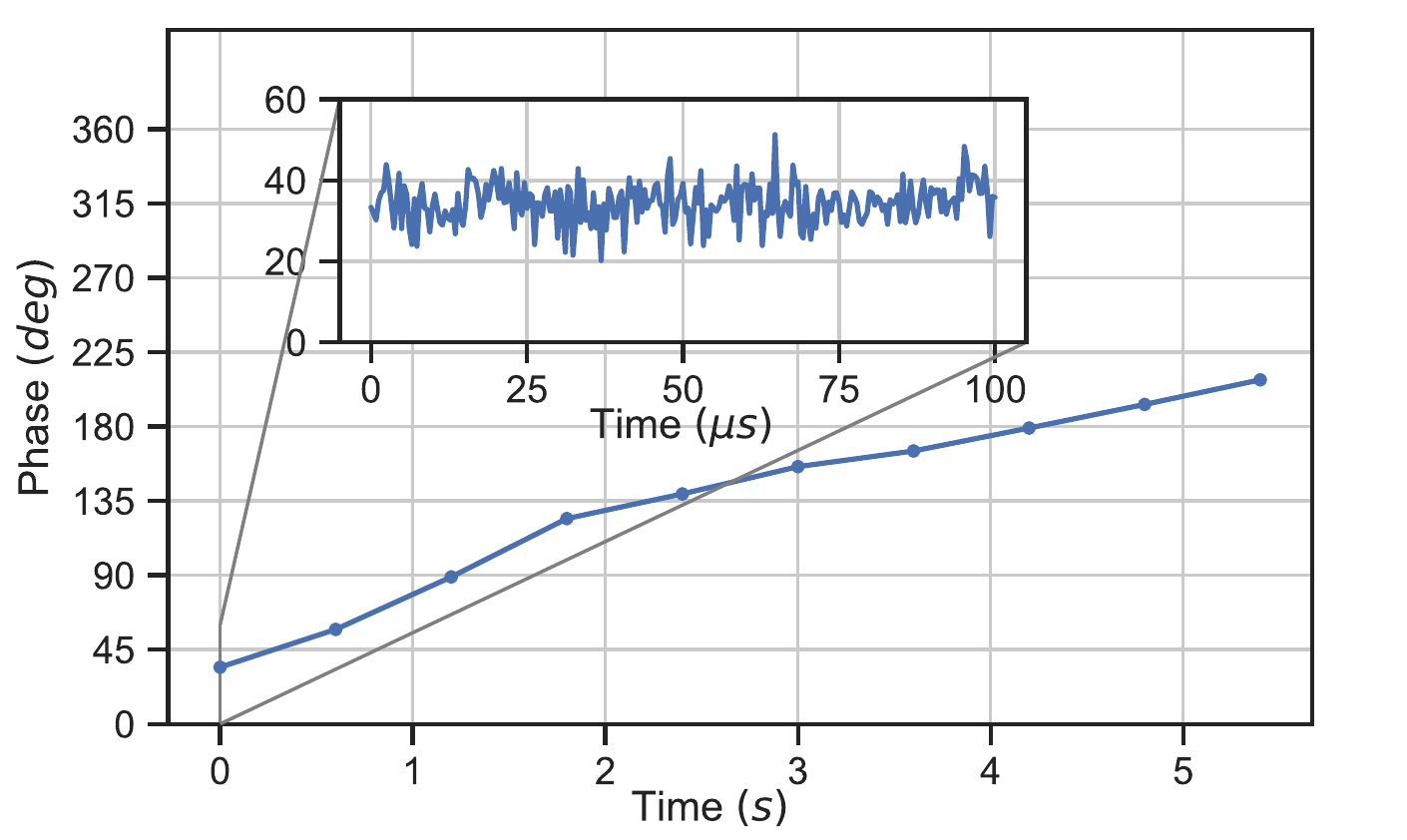}
\caption{Relative phase $\phi$ between signal and LO as a function of time. The system shows a drift of about $32\,\si{\deg\per\second}$, while for time scales comparable with the chunk size no drift is observed (see inset).}
\label{fig:phase}
\end{figure}

\subsection{Post-Processing and Software Analysis}
Since the signal and LO travel in different fibers, and no active phase stabilization is present, the relative phase between the signal and the LO is subjected to drifts over time. This means the two Gaussian distributions relative to $|\alpha \rangle $, and $|-\alpha \rangle$ will start to rotate around the center of the phase space, keeping fixed their overlap and their distance from the center. In order to compensate for these drifts, we perform a software phase tracking and compensation during the post-processing of the data. This procedure allows for a simpler experimental setup and is able to compensate fast phase fluctuations. In particular, during the post-processing we divide the acquired signal into ``chunks'' of
$n=1000$ samples. 
For each chunk, we calculate the centroids
of the measurement distribution in the phase space for each input state.
Choosing $n=1000$ is a good trade-off between the accuracy of centroid evaluation and phase stability. Indeed, each chunk lasts for less than $1\mu s$ while phase fluctuation are at the $ms$ scale.

In this way, it is possible to determine the relative phase $\phi$ between the signal and the LO. 
In Fig. {\ref{fig:phase}} the measured phase $\phi$ is plotted as a function of time.
Finally, we use the estimated $\phi$ to apply a linear classifier and assign an outcome $b$ to each measurement in the considered ``chunk''.

 After performing the procedure for all the chunks, then the experimental conditional probabilities $p(b|x)$ are estimated as 
 $\tilde p(b|x) =\frac {n_{b,x}} {n_x}$, where $n_{b,x}$ are the number of events for an output $b$ conditioned on an input $x$ and $n_x$ is the total number of transmitted states $x$.
The finite-size effects due to the limited statistics used to estimate the $\tilde p(b|x)$ can be included in the min-entropy bounds, as described in \cite{Brask2017}. In particular, we can use the Cheronoff-Hoeffding inequality \cite{Hoeffding1963} to build a confidence interval for the experimental probabilities $[\tilde p(b|x)-\Delta(\epsilon,n_x),\tilde p(b|x)+\Delta(\epsilon,n_x)]$, where $\epsilon$ is the security parameter and:
 \beq
 \Delta(\epsilon,n_x)=\sqrt{\frac{-\log_2(\epsilon)}{2n_x}}
 \eeq
This correction can be included in the objective function of the dual SDP, in order to obtain a reliable lower-bound on the min-entropy.

Finally, genuine random bits are extracted from the raw bit-string using a strong randomness extractor based on two-universal hashing functions implemented with Toeplitz matrices \cite{Tomamichel2011}. The security parameter $\epsilon_{RE}$ depends on the actual size of the matrices, but is always chosen $\epsilon_{RE}\leq10^{-10}$.

\section{Results}
We performed the experiment for different values of $\mu$.  
In each run of the protocol the mean photon number is estimated by the optical power captured via the calibrated PM in the signal branch. Given the mean photon number $\mu$ together with the measured probabilities $p(b|x)$, the SDP allows to evaluate the conditional min-entropy. 

\begin{figure}[t!]
\centering
\includegraphics[width=\linewidth]{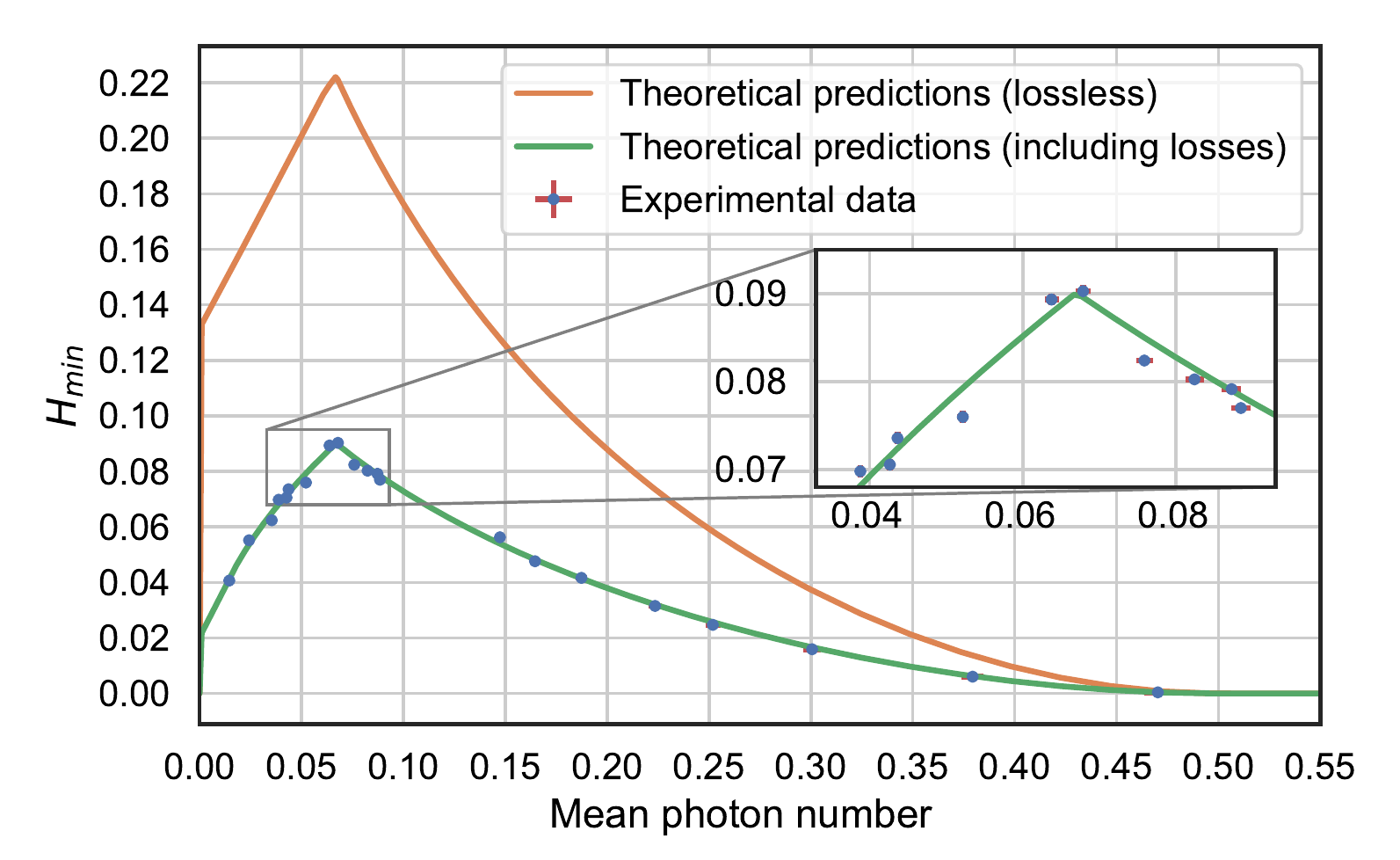}
\caption{Conditional min-entropy as a function of the mean photon number.  The orange curve is the numerical predictions obtained by SDP. The green one shows the numerical results of SDP when inefficiencies are considered and shows good agreement with experimental data (blue points).}
\label{fig:res}
\end{figure}

The experimentally estimated min-entropy and the corresponding theoretical values are displayed in Fig. \ref{fig:res}. For ideal detectors and lossless transmission (blue curve in Fig.~\ref{fig:res}), the predicted maximum of  $H_{min}(b|x)$ is $0.23$ bits per measurement, while experimentally, we achieved $0.09$ bits per measurement, 
corresponding to an absolute generation rate of about 113 Mbps.
This result can be explained by including in the model losses and detector's inefficiency. In any practical implementation all these effects are inevitable and contribute to worse discrimination of the incoming states decreasing the number of secure random bits with respect to the ideal detector case. 

In order to understand what is the significance of these non-idealities, we fitted the experimental probabilities to a model where noiseless but inefficient detectors (with efficiency $\eta$) are used for the heterodyne detection. A single value of the parameter $\eta$ is used to obtain the entire distribution of the min-entropy as a function of $\alpha$. The MLE fit returns a value of $\eta= 0.173\pm 0.002$. 
With this value injected in our theoretical model, we rerun the SDP, and we observe that experimental data and theoretical predictions are in perfect agreement.
Finally, a statistical randomness test is performed to check for possible problems and patterns  in the generated bit string. To do so, we run the NIST and ENT batteries of tests on the extracted numbers. In both cases, all the tests were passed; however, these tests don’t certify the randomness but at least reveals possible patterns due to classical noises. In short, it is a way to double-check that our system is operating as expected.

\section{Conclusion}
In conclusion, we realized a simple Semi-DI QRNG solution, based on heterodyne detection and a single assumption on the maximal energy of the prepared quantum states \cite{Himbeeck2017semidevice}.
With respect to the overlap bound introduced in \cite{Brask2017}, this assumption is easier to check experimentally and in this work is carefully monitored in real-time.
From the experimental point of view, our realization is based on the prepare-and-measure scenario implemented in a simple all-in-fiber optical setup with only COTS components. 
Our setup exploits heterodyne detection, as it provides several key advantages respect to other measurement strategies. First, it allows us to use commercial high-speed balanced detectors instead of slow and expensive single photon detectors, greatly increasing the performances while reducing the experimental complexity of the system.
Secondly, by sampling the entire phase space, it allows us to track the unavoidable phase drift between the signals and the LO. In this way, fast drifts can be compensated via software during the post-processing, avoiding the need of a complex active phase stabilization system.
With this scheme, we are able to generate and certify private random bits at a rate higher than $113$Mbps.

To conclude, we believe that our QRNG represents a great trade-off between the trust on the device, ease-of-implementation, and performance making it an attractive solution for many practical applications.

\acknowledgements

We thank Davide Rusca for useful discussion.

This work was supported by: ``Fondazione Cassa di Risparmio di Padova e Rovigo'' with the project QUASAR funded within the call ``Ricerca Scientifica di Eccellenza 2018''; Italian Space Agency with the project ``Realizzazione integrata di un Generatore Quantistico di Numeri Casuali-QRNG''; MIUR (Italian Minister for Education) under the initiative ``Departments of Excellence'' (Law 232/2016); EU-H2020 program under the Marie Sklodowska Curie action, project QCALL (Grant No. GA 675662).


\newpage
\appendix

\section{Semi-Definite Programming}
\label{sec:SDI}
In this section we will provide additional information regarding the bound on the guessing probability presented in the main text and its formulation via Semidefinite Programming (SDP) \cite{boyd2004convex}. 

If we consider the case of balanced inputs $p(x)=\frac{1}{2}$ and a fixed overlap $\Lambda=1-2\mu$, then for given measured probabilities $p(b|x)$, the guessing probability of an adversary $P_g$ can be upper-bounded by

\beq
\begin{aligned}
{P_g} &\le \frac{1}{2}\mathop{\max }\limits_{\{p_\lambda ,\rho_x^\lambda,\Pi _b^\lambda\}} \sum\limits_{x = 0}^1
{\sum_\lambda {{p_\lambda }\max \{ \tr[\rho _x^\lambda \Pi _0^\lambda ],\tr[\rho _x^\lambda \Pi _1^\lambda ]}\} } 
\end{aligned}
\label{eq:pguess_initial}
\eeq
where the maximization over $\{p_\lambda ,\rho_x^\lambda,\Pi _b^\lambda\}$ is performed subjected to the two constraints given in 
eqs. \eqref{eq:overlap_bound} and
\eqref{eq:constraints}, corresponding to the overlap bound
 and the compatibility with the experimental probabilities $p(b|x)$.
 In the previous equation $\lambda$ represents classical shared randomness between preparation and measurement device, $p_\lambda=p(\lambda)$ is the associated probability, $\rho_x^\lambda = |\psi_x^\lambda \rangle \langle \psi_x^\lambda|$ are the density matrices of the prepared states and $\{\Pi _b^\lambda\}$ are a binary outcome positive-operator valued measurement  (POVM) modeling the measurement device.

Unfortunately in this form there is no efficient way to compute the bound.
However, following the same approach described in \cite{Brask2017} and considering that, due to unitary invariance of \eqref{eq:pguess_initial}, the $\ket{\psi_x^\lambda}$ can be written without loss of generality as $\ket{\psi_0}=\ket{0}$, $\ket{\psi_1}=\Lambda\ket{0}+\sqrt{1-\Lambda^2}\ket{1}$ with $\ket0$ and $\ket1$ orthogonal states,
it is possible to express the problem as an SDP.

In particular, the problem can be rewritten as:
\begin{maxi}|l|
  {
  M_b^{\lambda_0,\lambda_1}
  }
  {
  \frac{1}{2}\sum\limits_{x = 0}^1 
  \sum\limits_{{\lambda _0},{\lambda _1} = 0}^1 \tr[{\rho _x}
  M_{\lambda_x}^{{\lambda _0},{\lambda _1}}]
  }
  {}{}
  \addConstraint{M_b^{{\lambda _0},{\lambda _1}} = {(M_b^{{\lambda _0},{\lambda _1}})^\dag }}
  \addConstraint{M_b^{{\lambda _0},{\lambda _1}} \ge 0}
  \addConstraint{\sum\limits_b {M_b^{{\lambda _0},{\lambda _1}}}  = \frac12 \tr[\sum\limits_b {M_b^{{\lambda _0},{\lambda _1}}} ]\mathbb{I}}
  \addConstraint{\sum\limits_{{\lambda _0},{\lambda _1} = 0}^{1} {\tr[{\rho _x}
  M_b^{{\lambda _0},{\lambda _1}}]}  = p(b|x)}
  \label{eq:SDP_primal}
 \end{maxi}
 where, $ M_b^{{\lambda _0},{\lambda _1}} = {p_{{\lambda _0},{\lambda _1}}} \Pi _b^{{\lambda _0},{\lambda _1}} $.
This optimization problem defines the primal SDP, which can be efficiently solved numerically.

However, since the primal involves a maximization, a solution to the problem provides a lower-bound on the guessing probability $P_g$, not an upper-bound. Thus, if the solver doesn't converge to the exact solution it will over-estimate the true amount of private randomness.

This problem can be solved by considering the dual formulation of \ref{eq:SDP_primal}, which involves a minimization problem and returns an upper-bound on $P_g$.
Also in this case it is possible to follow the procedure described in \cite{Brask2017} to derive the dual problem, which can be written as:
\begin{mini}|l|
  {H^{\lambda _0,\lambda _1},\nu _{bx}}{ \hskip-0.3cm- \sum\limits_{b,x} {{\nu _{bx}}p(b|x)  } \hspace{3cm}   }{}{}
  \addConstraint{H^{{\lambda _0},{\lambda _1}}} = {\rm{ }}{({H^{{\lambda _0},{\lambda _1}}})^\dag}
  \addConstraint{\begin{aligned}
\sum\limits_x & {{\rho _x}(\frac{1}{2}{\delta _{{\lambda _{x,0}}}}{\delta _{b,0}} + \frac{1}{2}{\delta _{{\lambda _{x,1}}}}
\delta _{b,1} + {\nu _{bx}})} \\
&+ {H^{{\lambda _0},{\lambda _1}}} - \frac{1}{2}\tr[{H^{{\lambda _0},{\lambda _1}}}]\mathbb{I} \le 0
\end{aligned}}
  \label{eq:SDP_dual}
\end{mini}
where $\delta_{i,j}$ is the Kronecker delta.

Interestingly, this dual formulation provides other two advantages when compared to the primal. The first advantage is related to the speed of the computation. With the primal, every time we obtain new data $p(b|x)$, it is necessary to run again the SDP to obtain a bound on the guessing probabiliy $P_g$. With the dual, since the objective function is a linear function of the $p(b|x)$, after an optimal solution has been found, a new (sub-optimal) upper-bound can be easily evaluated for different $p(b|x)$, without running again the optimization. This aspect is particularly interesting for real-time applications, where the post-processing can be done efficiently using a Lookup table.
Finally, with the dual formulation finite-size effects due to the limited statistics can be easily taken into account.
If $\Delta(\epsilon,n)$ is the finite-size correction, calculated for example with a tail inequality such as the Cheronoff-Hoeffding \cite{Hoeffding1963}, then by using 
\beq
- \sum\limits_{b,x} \left( \nu _{bx}p(b|x) +|\nu_{b,x}|\Delta(\epsilon,n) \right)   
\eeq
as objective function in Eq. \eqref{eq:SDP_dual}, it is possible to obtain a reliable upper-bound.

\section{Heterodyne versus Homodyne Detection}
\label{sec:homo}
In this section, we show the differences between heterodyne and homodyne detection.

Homodyne detector measures along only one quadrature, such as $X$ or $P$, as it is represented in Fig.~\ref{fig:overlap}(A), and the information about the other quadrature is lost, see Fig.~\ref{fig:overlap}(B). 
Distinguishability between the two input states is reduced for values of $\theta$ (the phase between the states $\ket{\pm\alpha}$ and the LO) different from 0. Real-time phase stabilization is thus required for homodyne measurement.
Indeed, as shown in Fig.~\ref{fig:overlap}, when the phase is  $\theta=0$, the possibility of distinguishing the two input state by measuring the $X$ quadrature is maximum. As long as  $\theta \neq 0,\pi$, the distinguishability decreases. 

Subsequently, the conditional probabilities $p(b|x)$ of obtaining $b$ given $x$ varies as a function of $\theta$ and the conditional min-entropy changes accordingly.
Indeed, if the input states are the coherent states $\ket{\pm\alpha}$ with $\alpha=|\alpha|e^{i\theta}$, the predicted conditional probabilities with homodyne measurement over the $X$ quadrature are given by 
\beq
\begin{aligned}
p(0|0) = p(1|1) &= \frac{1}{2}(1 + \erf(\sqrt 2 |\alpha\cos\theta|))
\\
p(1|0) = p(0|1) &= \frac{1}{2}(1 - \erf(\sqrt 2 |\alpha\cos\theta|))
\end{aligned}
\tag{A.4}
\eeq
We note that, differently from \eqref{eq:conditional_prob}, the previous probabilities depend on $\theta$. Moreover, it is worth to note that the above probabilities coincide with \eqref{eq:conditional_prob}
for $\theta=\pi/4$.

On the other hand, information about both quadratures 
$X$ and $P$ are accessible by performing heterodyne detection, see Fig.~\ref{fig:overlap}(C).
Thus, the distinguishability is constant as a function of $\theta$ and the conditional probabilities $p(b|x)$ will not change.

\begin{figure}[h!]
\centering
\includegraphics[width=\linewidth]{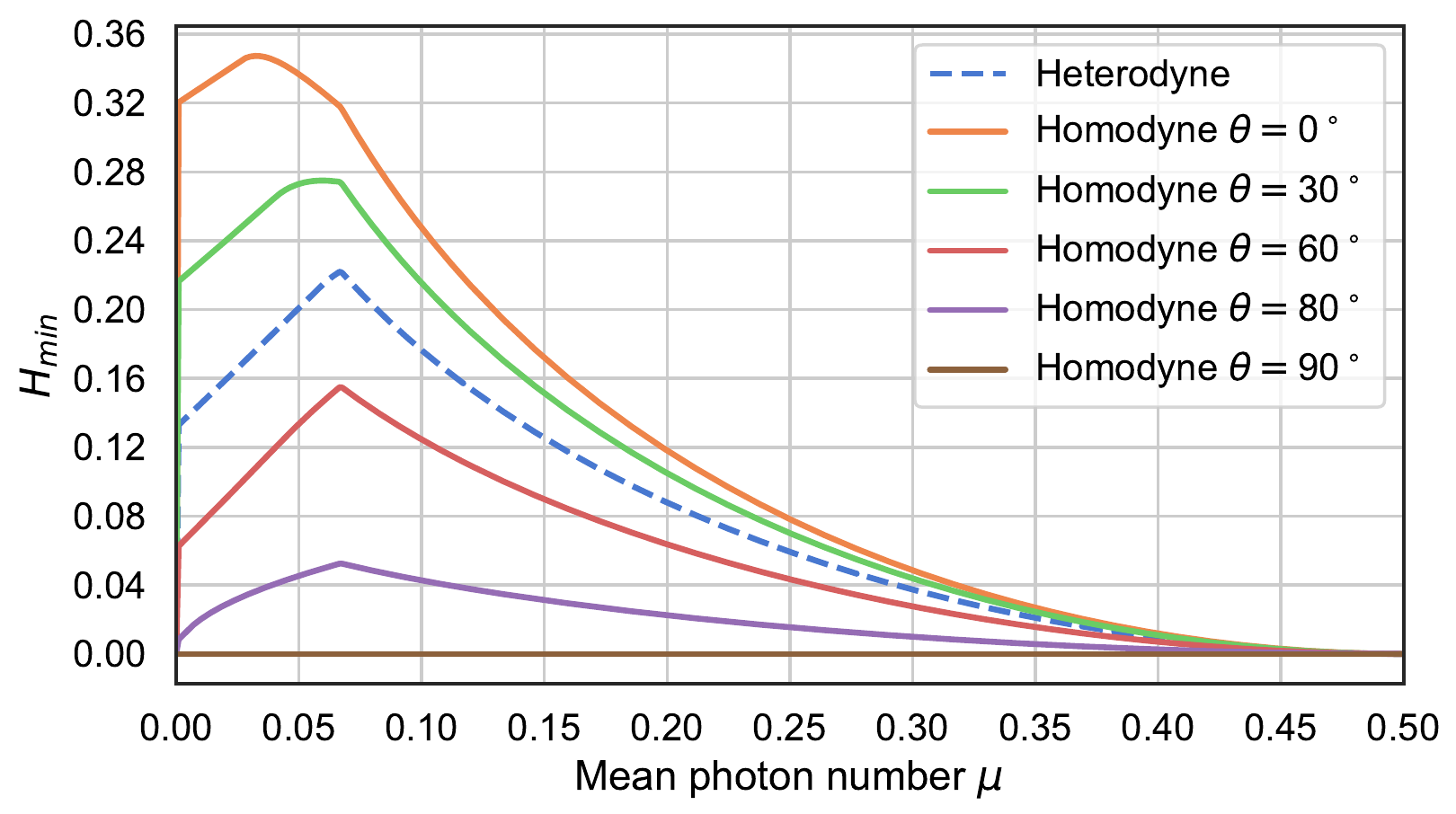}
\caption{In this figure, the conditional min-entropy is plotted as a function of the mean photon number for homodyne and heterodyne detection with different phases. It becomes clear that, when $\theta=90^{\circ}$, the min-entropy goes to zero, consequently no random bit can be extracted. This happens due to the fact that when $\theta=90^{\circ}$, the two input states show the same distribution in the $X$ quadrature, as if only one state was sent.}
\label{fig:homodyne}
\end{figure}
\begin{figure}[h!]
\centering
\includegraphics[width=\linewidth]{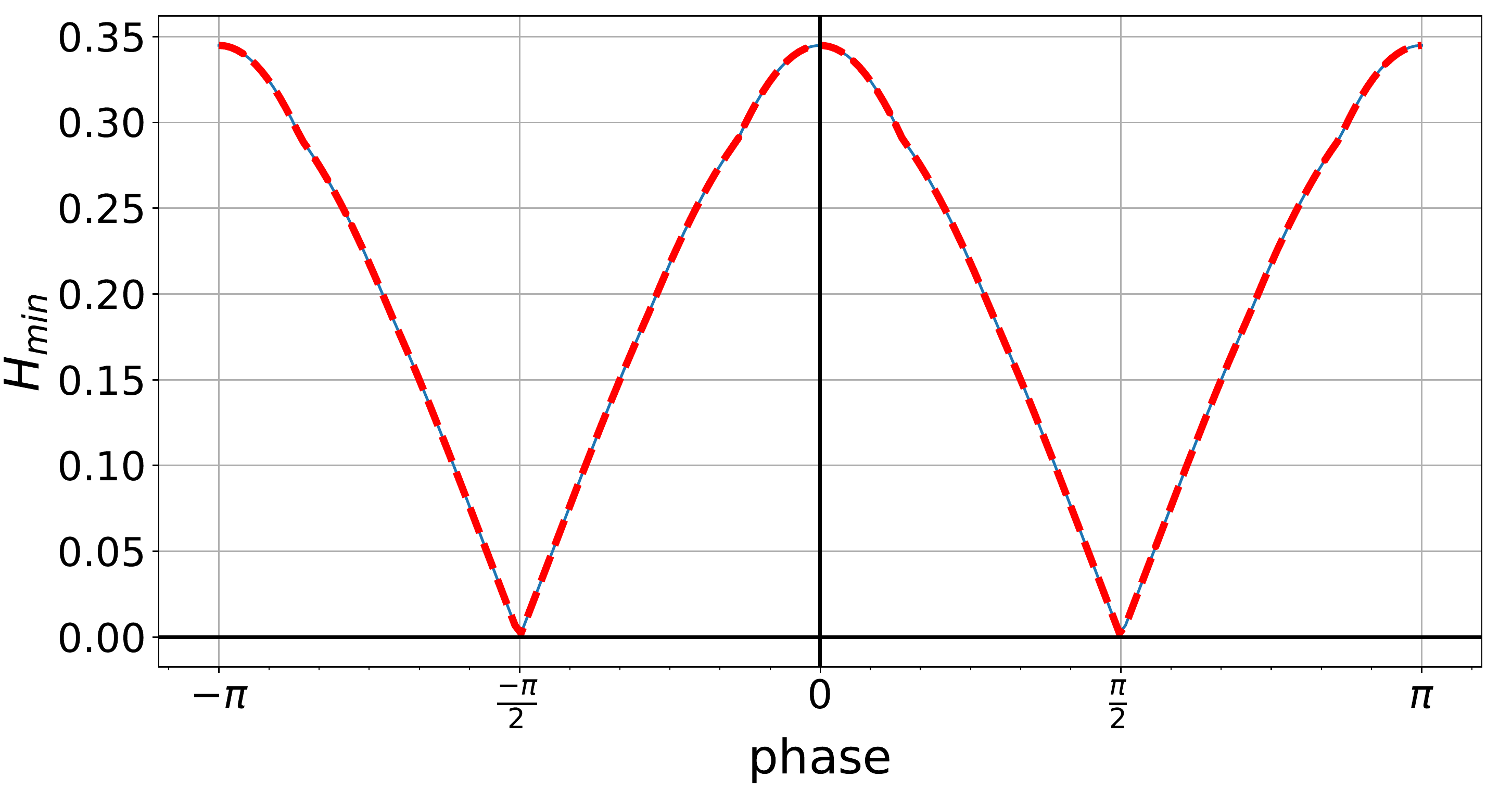}
\caption{In this figure, the conditional min-entropy is plotted as a function of the phase $\theta$ for homodyne detection. For each value of $\theta$, we choose the $\mu$ value that maximize the min-entropy.}
\label{fig:phase_hom}
\end{figure}

We show in Fig. \ref{fig:homodyne} the min-entropy in function of $\mu$ for different values of $\theta$ for the homodyne and heterodyne measurement. 
As predicted by conditional probabilities, the min-entropy depends on $\theta$ for the heterodyne measurement and for $\frac\pi4+n\pi<\theta<\frac34\pi+n\pi$ (with integer $n$) it is always lower than the heterodyne min-entropy.
We also show in Fig. \ref{fig:phase_hom} the maximum achievable min-entropy for heterodyne measurement in function of $\theta$ (for each $\theta$ we choose the $\mu$ value that maximize the min-entropy).

\end{document}